\documentclass[fleqn,10pt]{wlscirep}
\usepackage[utf8]{inputenc}
\usepackage[T1]{fontenc}
\usepackage{lineno}

\newcommand{\beq}{\begin{equation}}
\newcommand{\eeq}{\end{equation}}
\newcommand{\bea}{\begin{eqnarray}}
\newcommand{\eea}{\end{eqnarray}}
\newcommand{\be}{\begin{equation}}
\newcommand{\ee}{\end{equation}}

\usepackage[normalem]{ulem}

\colorlet{mgreen}{green!50!black!50!}

\title{Variant-driven multi-wave pattern of COVID-19 via a Machine Learning analysis of spike protein mutations}

\author[1,+]{Adele de Hoffer}
\author[2,3,+]{Shahram Vatani}
\author[2,3,+]{Corentin Cot}
\author[2,3,*]{Giacomo Cacciapaglia}
\author[4,5]{Maria Luisa Chiusano}
\author[6]{Andrea Cimarelli}
\author[7,8]{Francesco Conventi}
\author[9]{Antonio Giannini}
\author[2,3]{Stefan Hohenegger}
\author[7,10,11,12,*]{Francesco Sannino}

\affil[1]{Politecnico di Torino, Corso Duca degli Abruzzi 24, 10129 Torino, Italy}
\affil[2]{Institut de Physique des 2 Infinis (IP2I), CNRS/IN2P3, UMR5822, 69622 Villeurbanne, France}
\affil[3]{Universit\' e de Lyon, Universit\' e Claude Bernard Lyon 1, 69001 Lyon, France}
\affil[4]{Department of Agricultural Sciences, Universit\`a degli Studi di Napoli Federico II, Portici, 80055 Italy}
\affil[5]{Department of Research Infrastructures for Marine Biological Resources (RIMAR), Stazione Zoologica ``Anton Dohrn'', 80121 Napoli, Italy}
\affil[6]{Centre International de Recherche en Infectiologie (CIRI), Univ Lyon, Inserm, U1111, Universit\'e Claude Bernard Lyon 1, CNRS, UMR5308, ENS de Lyon, 46 All\'ee d'Italie, 69007 Lyon, France}
\affil[7]{INFN sezione di Napoli, Complesso Universitario di Monte S. Angelo Edificio 6, via Cintia, 80126 Napoli, Italy}
\affil[8]{Dipartimento di Ingegneria Universit\`a degli studi di Napoli Parthenope, Centro Direzionale di Napoli, Isola C 4, lato Sud, 80143 Napoli, Italy}
\affil[9] {University of Science and Technology of China (USTC), No.96,  JinZhai Road, Baohe District, Hefei, Anhui, 230026, China}
\affil[10]{Dipartimento di Fisica E. Pancini, Universit\`a di Napoli Federico II, Complesso Universitario di Monte S. Angelo Edificio 6, via Cintia, 80126 Napoli, Italy}
\affil[11]{Scuola Superiore Meridionale, Largo S. Marcellino 10, 80138 Napoli, Italy}
\affil[12]{CP3-Origins \& the Danish Institute for Advanced Study, University of Southern Denmark, Campusvej 55, DK-5230 Odense, Denmark}

\affil[*]{g.cacciapaglia@ipnl.in2p3.fr, sannino@cp3.sdu.dk}

\affil[+]{these authors contributed equally to this work}

\begin{abstract}
\noindent
Never before such a vast amount of data, including genome sequencing, has been collected for any viral pandemic than for the current case of COVID-19. This offers the possibility to trace the virus evolution and to assess the role mutations play in its spread within the population, in real time. To this end, we focused on the Spike protein for its central role in mediating viral outbreak and replication in host cells. Employing the Levenshtein distance on the Spike protein sequences, we designed a machine learning algorithm yielding a temporal clustering of the available dataset. From this, we were able to identify and define emerging persistent variants that are in agreement with known evidences. 
Our novel algorithm allowed us to define persistent variants as chains that remain stable over time and to highlight emerging variants of epidemiological interest as branching events that occur over time. Hence, we determined the relationship and temporal connection between variants of interest and the ensuing passage to dominance of the current variants of concern.
Remarkably, the analysis and the relevant tools introduced in our work serve as an early warning for the emergence of new persistent variants once the associated cluster reaches 1\% of the time-binned sequence data. We validated our approach and its effectiveness on the onset of the Alpha variant of concern.  We further predict that the recently identified lineage AY.4.2 (`Delta plus') is causing a new emerging variant. Comparing our findings with the epidemiological data we demonstrated that each new wave is dominated by a new emerging variant, thus confirming the hypothesis of the existence of a strong correlation between the birth of variants and the pandemic multi-wave temporal pattern. 
The above allows us to introduce the epidemiology of variants that we described via the Mutation epidemiological Renormalisation Group (MeRG) framework.
\end{abstract}
\begin{document}

\flushbottom
\maketitle
\thispagestyle{empty}

\section*{Highlights}

\begin{itemize}
\item {\bf Objectives} To study the relation among Spike protein mutations, the emergence of relevant variants and the multi-wave pattern of the COVID-19 pandemic. 

\item {\bf Setting} Genomic sequencing of the SARS-CoV-2 Spike proteins in the UK nations (England, Scotland, Wales). Epidemiological data for the number of infections in the UK nations, South Africa, California and India. 

\item {\bf Methodology} We design a machine learning algorithm, based on the Levenshtein distance on the Spike protein sequences, that leads to a temporal clustering of the available dataset, from which we define emerging persistent variants. The above allows us to introduce the epidemiology of variants that we described via the Mutation epidemiological Renormalisation Group (MeRG) framework.

\item {\bf Results} We show that:
\begin{enumerate}
\item Our approach, based only on the Spike protein sequence, allows to efficiently identify the variants of concern (VoCs) and of interest (VoIs), as well as other emerging variants occurring during the diffusion of the virus.
\item Within our time-ordered chain analysis, a branching relation emerges, thus permitting to reconstruct the evolutionary diversification of Spike variants and the establishment of the epidemiologically relevant ones.
\item Our analysis provides an early warning for the emergence of new persistent variants once its associated dominant Spike sequence reaches 1\% of the time-binned sequence data. Validation on the onset of the Alpha VoC shows that our early warning is triggered 6 weeks before the WHO classification decision.
\item Comparison with the epidemiological data demonstrates that each new wave is dominated by a new emerging variant, thus confirming the hypothesis that there is a strong correlation between the emergence of variants and the multi-wave temporal pattern depicting the viral spread. 
\item A theory of variant epidemiology is established, which describes the temporal evolution of the number of infected by different emerging variants via the MeRG approach. This is corroborated by empirical data.
\end{enumerate}

\item {\bf Conclusions} Applying a ML approach to the temporal variability of the Spike protein sequence enables us to identify, classify and track emerging virus variants. Our analysis is unbiased, in the sense that it does not require any prior knowledge of the variant characteristics, and our results are validated by other informed methods that define variants based on the complete genome. Furthermore, correlating persistent variants of our approach to epidemiological data, we discover that each new wave of the COVID-19 pandemic is driven and dominated by a new emerging variant. Our results are therefore indispensable for further studies on the evolution of SARS-CoV-2 and the prediction of evolutionary patterns that determine current and future mutations of the Spike proteins, as well as their diversification and persistence during the viral spread. Moreover, our ML algorithm works as an efficient early warning system for the emergence of new persistent variants that may pose a threat of triggering a new wave of COVID-19. Capable of a timely identification of potential new epidemiological threats when the variant only represents 1\% of the new sequences, our ML strategy is a crucial tool for decision makers to define short and long term strategies to curb future outbreaks. The same methodology can be applied to other viral diseases,  influenza included, if sufficient sequencing data is available.

\end{itemize}

\vspace{1cm}

\section*{Introduction}
It is of primary importance to understand the diffusion of a virus and the establishment of its variants, to further understand the infection mechanisms and fight the associated disease, especially in view of an efficient vaccination campaign. 
This task could not be efficiently approached in past extended pandemics caused by infectious diseases, like for instance the ``Spanish'' Influenza of 1918-19 \cite{1918influenza}, mainly due to the paucity of the available data.
The current COVID-19 crisis is, on the contrary, revolutionising our understanding of pandemics because of the efficient collection of a large amount of data (e.g. genome sequencing, epidemiology etc) in real time, which allows for a timely identification of viral variants that successfully radiate throughout the world. 
Among the mutations that characterise SARS-CoV-2 variants, those that can be traced along the spike protein (S) sequence are major, although  not unique, drivers of viral spread in the human population for the role that this protein plays in mediating the virus entrance into target cells as well as for its role in mediating escape from antibody responses. 
In this work, a \emph{mutation} is a single change in the amino acid sequence of the Spike protein (substitution, addition, deletion), while a \emph{Spike variant}, or simply \emph{variant}, is a unique sequence of amino acids in the Spike protein that appears in clusters. 
 Like other coronaviruses, the SARS-CoV-2 has relatively low mutation rates \cite{Sanjuan9733}, nevertheless the current COVID-19 pandemic has seen the emergence of several epidemiologically relevant variants. Efficient nucleotide sequencing has allowed to track sequence mutations along the genome of SARS-CoV-2, and to identify dangerous variants \cite{Plante2020,KORBER2020812} that  appeared to increase the infectivity compared to the initial form that was sequenced from the outbreak in Wuhan, China~\cite{Wu2020sequence} (GenBank: MN908947.3).  Since the second half of 2020, variants of concern (VoCs) and of interest (VoIs) have been identified in various regions of the world. For instance, following the naming scheme of the WHO \cite{WHOvarname} (Pango lineage \cite{Pango}, GISAID \cite{GISAID1,GISAID2}):  The Alpha VoC (B.1.1.7, GRY), first identified in September 2020 in the UK \cite{britishVOC,Mahasem4944}; the Beta VoC (B.1.351, GH/501Y.V2) first found in South Africa in May 2020 \cite{Tegally2020.12.21.20248640}; the Gamma VoC (P.1, GR/501Y.V3) first detected in Brazil in November 2020\cite{SabinoBrazil}, which has been spreading in Manaus notwithstanding the high rate of previous infections; the Delta VoC (B.1.617.2, G/478K.V1) identified in India in October 2020; and the Epsilon VoI (B.1.427+429, GH/452R.V1) found in California in March 2020 \cite{Pater2021.01.11.426287}. An exhaustive list can be found on the WHO website (\href{https://www.who.int/en/activities/tracking-SARS-CoV-2-variants/}{www.who.int}). Considering the Alpha VoC as an example, it has been possible to study its infectious power in lab experiments, finding a higher rate of transmission by $67$--$75$\%, compared to the previous ones \cite{Mahasem4944}. The transmission advantage has been confirmed by  epidemiological data in the UK \cite{Rasigade2020,Volz2020.12.30.20249034}. Most analyses of the epidemiological data are done applying the time-honoured compartmental models of the SIR type \cite{Kermack:1927,PERC20171,WANG20151}, appropriately extended by including more compartments \cite{Giordano2021}. The main drawback in this approach is the large number of parameters, which need to be fixed by hand or extracted from the data.
In this work, we bypassed this bottleneck by using a simplified and effective approach based on theoretical physics methods, the epidemic Renormalisation Group (eRG) framework \cite{DellaMorte:2020wlc,Cacciapaglia:2020mjf,cacciapaglia2020second}, combined with information directly extracted from the Spike protein sequence via a simple Machine Learning approach. This novel method allowed us to analyse, at the same time, the variability of the SARS-CoV-2 Spike protein in multiple countries and regions of the world, and thus provide a direct comparison of the epidemiological impact of the different Spike variants. A theoretical analysis of the variants within the eRG framework is presented in a companion publication \cite{cacciapaglia2021MeRG}.

In the present work, we analysed the protein sequence data for the UK nations downloaded from the GISAID repository \cite{GISAID1,GISAID2}. We implemented a simple Machine Learning (ML) algorithm based on the Levenshtein measure (LM) \cite{Levenshtein,Levenshtein2} in order to cluster protein sequences based on their distance in terms of number of amino acid substitutions (i.e. the number of amino acid mutations needed to transform one sequence into the other, or vice-versa). The clusters have been defined by setting cutoffs on the Ward distance between branches of a proximity tree, built by the use of a standard hierarchical clustering algorithm. We applied the clustering algorithm on the data binned in temporal units, specifically here identified by months or weeks. Each time unit may include more clusters according to the cutoffs set on the Ward distance and, within each cluster, the dominant variant is the most frequent in terms of identical sequences over the total number of sequences in the cluster. We then developed an algorithm that links clusters appearing in consecutive time units and creates chains of clusters that share the same dominant Spike variant. Empirically we determined that chains that persist for longer than three time units identify emerging variants. In order to reconstruct the origin of each emerging variant, we associated the initial cluster of each chain with a cluster from the previous time unit that maximises the overlap in their sequence content. Hence, a branching relation emerges in our procedure. For clarity, in this work we use the following definitions for variants and mutations:
\begin{enumerate}
\item A \emph{dominant variant} in a cluster is the Spike variant that is most frequently appearing in the cluster. Note that the chains are created by linking consecutive clusters when they possess the same dominant variant. Subdominant variants in each cluster are retained if their frequency is above 1\% in the cluster.
\item An \emph{emerging variant} is defined as an established chain that contain more than three consecutive clusters, defined using our linkage algorithm. This criterium is established empirically from the results of the chain reconstruction.
\end{enumerate}
It should be noted that some of the emerging variants defined by our procedure can be associated to VoCs and VoIs, as defined by the WHO, as they share the same characteristic Spike mutations.

The procedure above has been independently repeated for each geographical region in our study.  
We validated our results by showing that our approach identifies the Alpha VoC, independently, in all the distinct UK regions we studied. Once the dominant variants were identified, we analysed their temporal spreading within the affected population. Given that only a small fraction of the infected individuals have their viral charge sampled and sequenced, we estimated the number of people infected by each variant by multiplying the number of positive tests by the rate of occurrence of each variant in the sequencing data. This rough approximation allows us to reliably extract the temporal evolution of each variant in the population. Note that each infected individual is, in practice, associated to the variant that is most frequently reappearing in their viral charge, following the practice of the sequence reporting. Thus, the data we use track the time development of the dominance of each variant at the individual level.

To analyse the time evolution of the individuals infected by each variant, we employed the economical eRG approach \cite{DellaMorte:2020wlc} that allows to organise the pandemic waves according to temporal symmetry principles similar to those found in high energy physics \cite{Wilson:1971bg,Wilson:1971dh}. The approach has been extensively tested \cite{cacciapaglia2020second,cacciapaglia2020better}, confronted with traditional SIR compartmental models \cite{Della_Morte_2021}, and, last but not least, summarised in a comprehensive review alongside other approaches \cite{cacciapaglia2021field}.  
The economy of the model rests in the fact that, once the overall number of infected individuals is fixed,  the diffusion rate of the virus is influenced by  a single  parameter $\gamma$ that measures the speed at which the virus spreads in the population. This value can be extracted by fitting the number of new daily infections or the cumulated number of infections. We remark that this approach  can be put in correspondence to a SIR model with time-dependent reproduction number $R_0 (t)$ \cite{Della_Morte_2021}, which fits the data better than traditional compartmental models with constant parameters.
We applied the eRG to each variant, thus yielding a classification of their aggressiveness via a single quantifier: their individual $\gamma$.  
A visual summary of the methodology followed by our analysis with its main outcomes is shown in Fig.~\ref{fig:1}, while more details are reported in the supplementary material.

The main goal of this work is to understand the viral dynamics that characterises wave patterns stemming from infectious diseases like COVID-19. The eRG approach additionally offers a natural mathematical understanding in terms of the dynamical flow of the system \cite{cacciapaglia2020multiwave,cacciapaglia2020evidence}.
 Importantly, by employing ML analysis to genomic data, we discovered that each pandemic wave is driven by a single emerging and persisting variant. The findings demonstrate that the variant dynamics is one of the main engines behind the emergence of wave patterns for COVID-19. This result can be used as a template for similar infectious diseases. As direct consequence of our study we propose a novel evolutionary model for the interpretation of the virus diffusion that is mutation driven.

\begin{figure}[tb!]
\begin{center}
\includegraphics[width=.99 \textwidth]{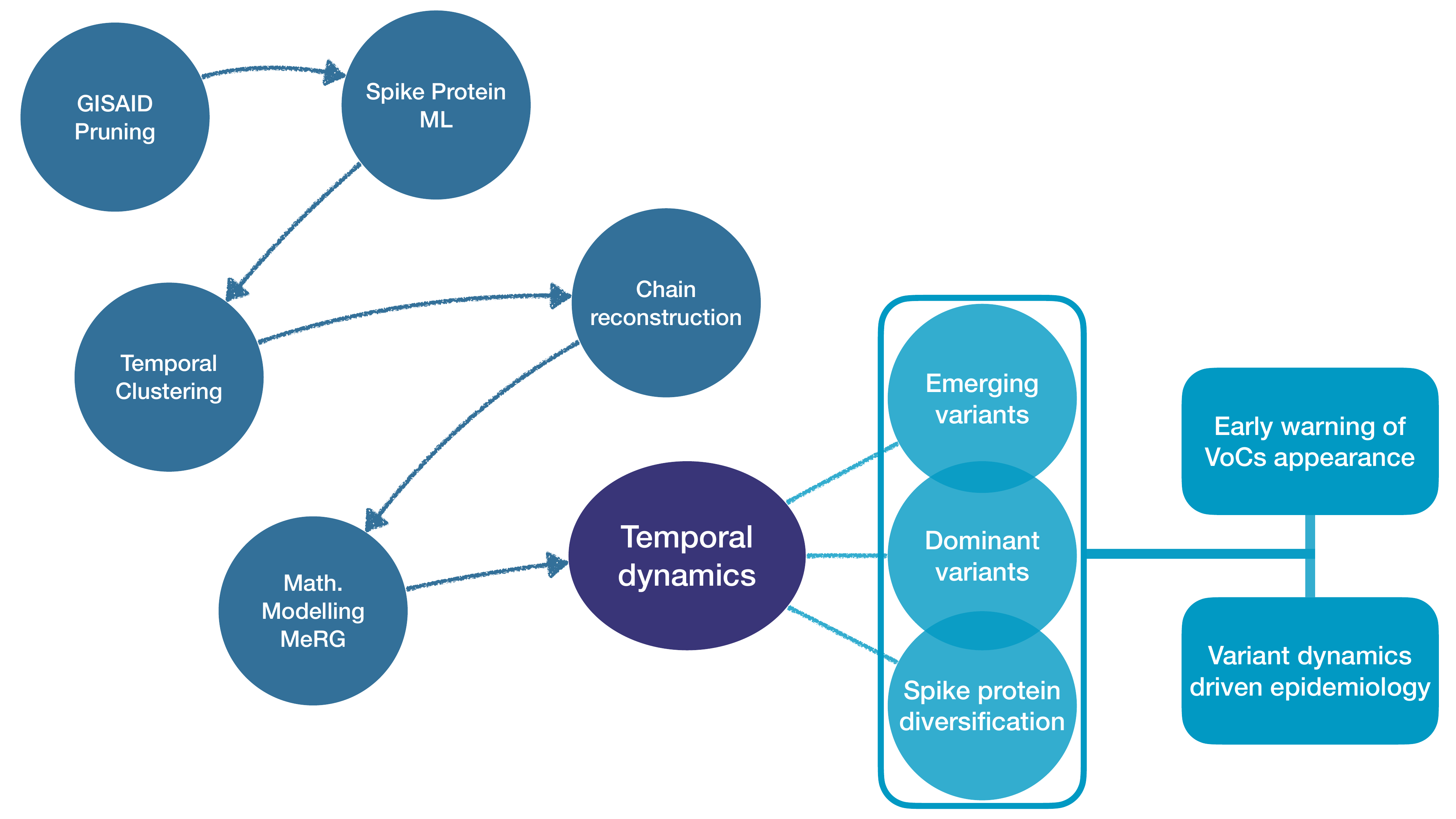}
\end{center}
\caption{{\bf Methodology and main outcomes.} Schematic representation of the work-flow we follow.}
\label{fig:1}
\end{figure}

\section*{Results}

Spike protein sequences have been extracted from the GISAID repository on a country-specific basis and the date-stamp associated to each sequence has been used to obtain a temporal dimension of viral variants appearance.  Note that each genome sequence in the GISAID data collection corresponds to the most frequent Spike variant occurring in a single infected individual. The pruned dataset (see supplementary material) has been clustered per month to obtain groups of distinct variants.  Firstly, we computed the LM between each pair of sequences, thus counting by an unweighted approach the minimal number of amino acid substitutions, deletions and insertions needed to transform one sequence into the other, and vice-versa. 
Secondly, the algorithm constructed a tree of proximity by pairing sequences that are the closest to each other into a branch. To combine branches that contain more than one sequence, we used Ward's method, after having checked that other choices do not significantly affect the results (more details in the supplementary material). The tree is completed when all sequences are grouped into a single branch. To define the clusters, we considered a cutoff in the distance so that branches whose Ward distance is larger than the cutoff are considered as separate clusters. We applied the same cutoff to all branches. 

As England has the largest available sequencing sample, with 646.697 sequences as of the end of August 2021, we mainly focused on this dataset. This  minimises statistical and sampling bias errors. After pruning,  461.122 sequences were retained, out of which we identified 13.887 distinct ones.

\begin{figure}[tb!]
\begin{center}
\includegraphics[width=.95 \textwidth]{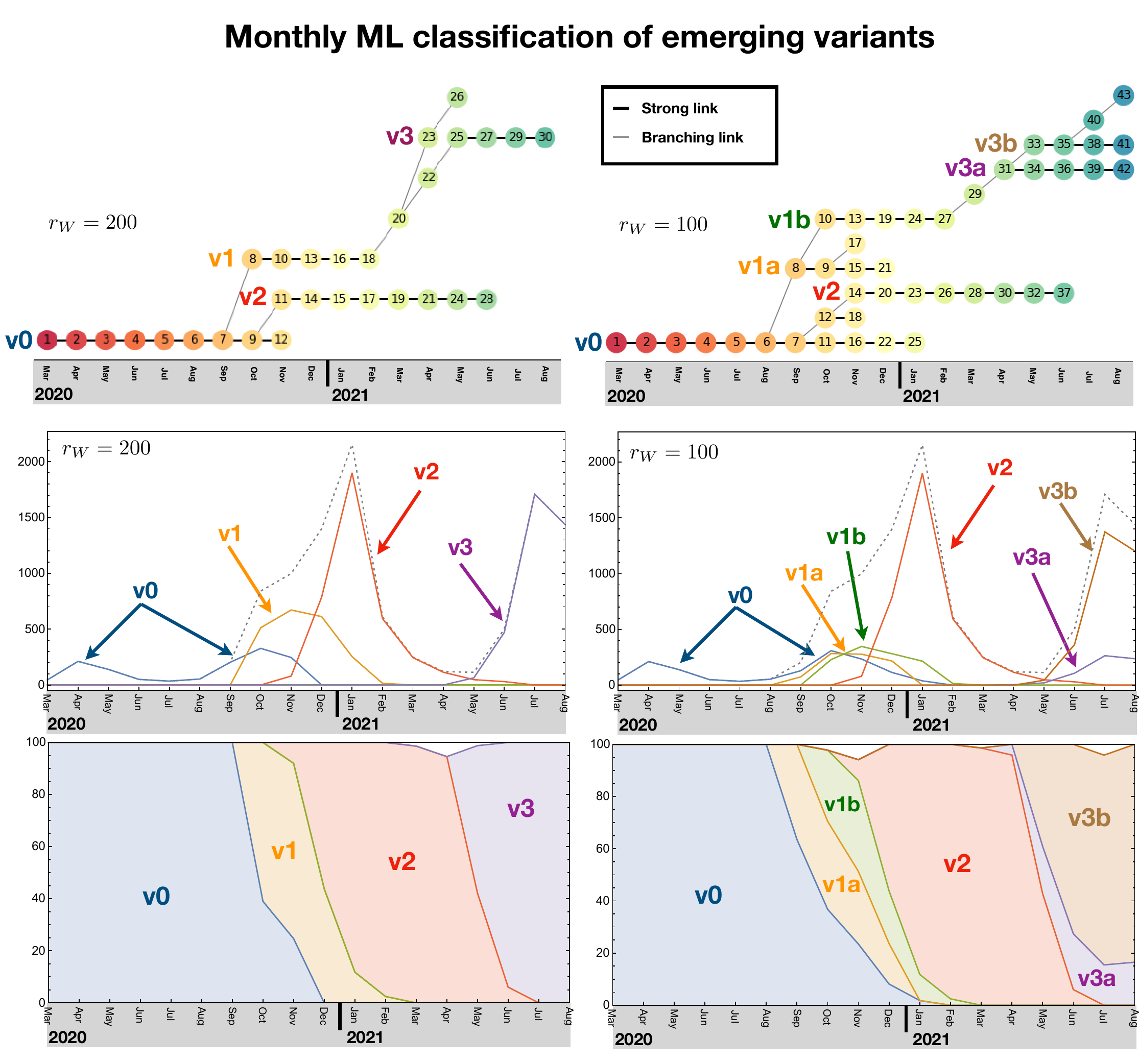}
\end{center}
\caption{{\bf Monthly ML analysis and chain variants.} The clusters are linked to form chains, which are then identified with emerging variants, as shown in the top plots. In the middle and bottom plots we show the number of monthly infected per 100k inhabitants and percentage of occurrence for each emerging variant.  The left plots correspond to a cutoff in the Ward distance of $r_W = 100$ while the right ones to $r_W = 200$. Note that the chains v2 and v3 for $r_W = 200$ can be associated to the Alpha and Delta VoC, respectively.}
\label{fig:4}
\end{figure}

\subsubsection*{Emerging variants as time-ordered cluster chains}

The time evolution and emergence of SARS-CoV-2 variants can be studied by applying our ML algorithm to the Spike sequence data binned in time, by calendar month.  Hence, we have divided the sequence dataset for England following the date tag in the GISAID repository. For each month, we run the ML algorithm on the pruned data to define clusters, retaining only the ones comprising at least 1\% of the monthly dataset. 
The cutoff on the Ward distance $r_W$ between branches, as well as the 1\% threshold above, were chosen to optimise the coverage of the dataset (i.e. we required that the defined clusters cover at least 90\% of the data) while  keeping the number of clusters below 10. The optimisation analysis, presented in the supplementary material, showed that the optimal range for the branch cutoff is $r_W \in [50, 200]$.
After this, we compared the clusters in consecutive months to link those with a ``strong similarity'', i.e. those that share the same dominant sequence (strong links). More details on this procedure and its validation can be found in the supplementary material.
The linkage algorithm we employed allowed to define  ``chains of clusters'' that we associate to emerging variants. The results are shown in Fig.~\ref{fig:4} for two choices of the Ward distance: $r_W = 100$ in the left and $r_W = 200$ in the right plots.  
For the two choices, we identified 6 and 4 cluster chains, respectively, that last more than three months. In the middle and bottom rows of Fig.~\ref{fig:4} we show the new monthly infections (per 100k inhabitants) and the frequencies of the cluster chains, which we identify as emerging variants in the following. In this respect, the results for $r_W=200$ can be directly compared to the VoCs identified by the WHO: Comparing the frequencies of occurrence, we see that v2 can be associated to the Alpha VoC, while v3 matches the epidemiological data for the Delta VoC. We also checked that the dominant Spike variant for the two chains presents the mutations characteristic of the two VoCs: N501Y, D614Y and P681H for the Alpha VoC; L452R, T478K, D614G and P681R for the Delta VoC.  
These results have been corroborated by a cluster analysis of the global dataset, without time binning, and by a similar analysis for the data of Wales and Scotland, as shown in the supplementary material.

The chain analysis, however, allowed us to better probe the time evolution and emergence of the persisting variants. To do so, for the clusters at the beginning of each chain, we defined a branching link with the cluster in the previous month. These connections are shown as grey diagonal links in the top plots of Fig.~\ref{fig:4}. 
From the case $r_W = 200$, we clearly see that v1, which is responsible for the second wave, branched off from v0 in October 2020. Similarly, v2, which corresponds to the Alpha VoC, also branched off from v0 a month later. The Delta VoC v3, instead, developed from v1 from February to May 2021, via two intermediate clusters, 20 and 22. Finally we see the emergence of a branch, 20--23--26, which died off being dominated by the Delta VoC starting with cluster 25. 
By lowering the cutoff that defines clusters, see left plots in Fig.~\ref{fig:4} for $r_W = 100$, one can see how v1 splits in two distinct, but closely related, chains, as well as the Delta VoC v3. The Delta VoC is now seen as branching off from v1b. The closeness of the clusters splitting from v1 and v3 is confirmed by comparing the dominant Spike sequences, showing that v1b differs from v1a only by the mutation L18F, while v3b differs from v3a only by the mutation T95I.
In particular, cluster 43 emerged in August 2021 and its dominant variant bears the Y145H and A222V mutations that identify the AY.4.2 lineage (`Delta plus' variant) \cite{AY.4.2}, which has been classified by Pango \cite{Pango} at the beginning of September. Out of the many new lineages that have been recently isolated, only this one is highlighted by our ML analysis.  As such, and with the caveat that our analysis includes only data up to August 2021, the ability of this novel variant to give raise to a stable chain in the near future deserves close attention.

These results firstly show that the phylogenetic relation between variants emerges from our simple ML algorithm applied exclusively to the Spike protein sequence. Furthermore, we see a distinctive pattern relating the emergence of a persistent variant and the exponential increase in infections that ignites a new pandemic wave. A new wave only emerges when a new variant is generated, which has the virological strength to overcome the old ones. This is seen very clearly with v2 (or Alpha VoC) which spins off from v0 closely to v1 and takes over by generating a third wave. We also see the emergence of short-lived variants that do not have the power to start a new wave and therefore die off without infecting a sizeable number of individuals.
All short-lived chains have less than two clusters, hence we define a minimum length of three for persisting chains.

\begin{figure}[tb!]
\begin{center}
\includegraphics[width=.99 \textwidth]{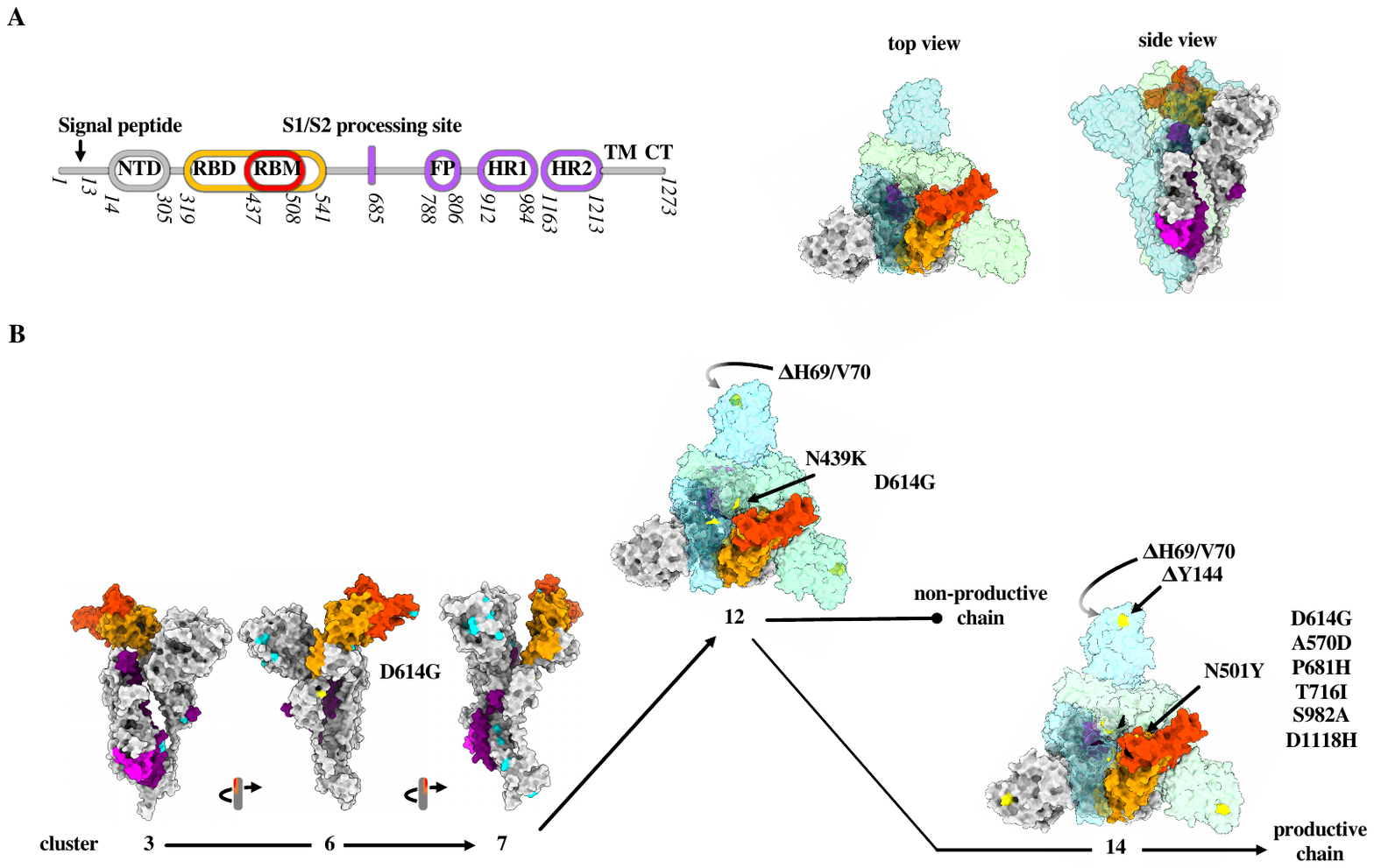}
\end{center}
\caption{{\bf Temporal analysis of mutations arising in the Spike protein during the genesis of the Alpha VoC.} A) Schematic representation of the SARS-CoV-2 Spike protein (S): N terminal domain (NTD); receptor binding domain and motif (RBD and RBM, respectively); fusion peptide (FP); heptad repeat 1 and 2 (HR1 and HR2, respectively); transmembrane domain (TM) and cytoplasmic tail (CT). Top and side view of the trimeric Spike protein in its closed conformation \cite{Wrobel2020} (PDB: 6ZGI). For simplicity, the colour codes of the different domains are provided for a single chain of the trimer. B) Accumulation of the different mutations in the different clusters leading to the establishment of the Alpha VoC: domains colour codes as in A) for a single S; yellow indicates mutations fixed in the dominant variant and cyan indicates mutations appearing in subdominant ones. In cluster 14 ($r_W = 100$), the original position of the N439K mutation that was lost to the profit of the near N501Y mutation is marked in black.}
\label{fig:structure}
\end{figure}

\subsubsection*{Features of Spike mutation dynamics within stable chains}

This method allows for a temporal analysis of the accumulation of viral Spike variants at different resolution, according to the chosen $r_w$. The $r_w=100$ analysis, for example, highlights how the Alpha variant (v2 stable chain comprised between clusters 14 and 37) arouse from cluster 7 (v0 stable chain of clusters 1 to 25) through an ephemeral intermediate cluster (cluster 12). The mapping of mutations that appeared in both dominant and subdominant variants in each cluster over time on the cryo-EM trimeric structure of the Spike protein \cite{Wrobel2020} is of interest (Fig.~\ref{fig:structure}A and \ref{fig:structure}B). In addition to the mutations fixed in the dominant Spike variant, mutations in subdominant ones (detected at equal/above 1\% of the entire pool of sequences in each cluster, and marked by cyan dots in the 3D spike models in Fig.~\ref{fig:structure}) accumulate with increasing frequency along each chain. This is not surprising per se and it likely reflects high viral replication rates over time at the population level. While so far mutations that become fixed in the dominant variant appear and reach dominance within 1 month, the fixation of the H69/V70 deletion seems to have undergone a transient sampling state with the N439K mutation in the RBM.  Viruses bearing the N439K Spike mutation have been characterised ex vivo and the mutation had been described to allow for antibody-mediated immunity escape while not affecting viral fitness \cite{Thomson2021}. Given that this mutation rapidly vanished (non-productive chain 12-18), it is clear that such mutation exhibited a defect in viral fitness that cannot be recapitulated with ex vivo studies. However, the N439K mutation may still have served an important role in the emergence of the subsequent dominant variant by allowing sufficient time for variants bearing the H69/V70 deletion to combine with more advantageous mutations (N501Y, DY144 etc).

\begin{figure}[tb!]
\begin{center}
\includegraphics[width=.99 \textwidth]{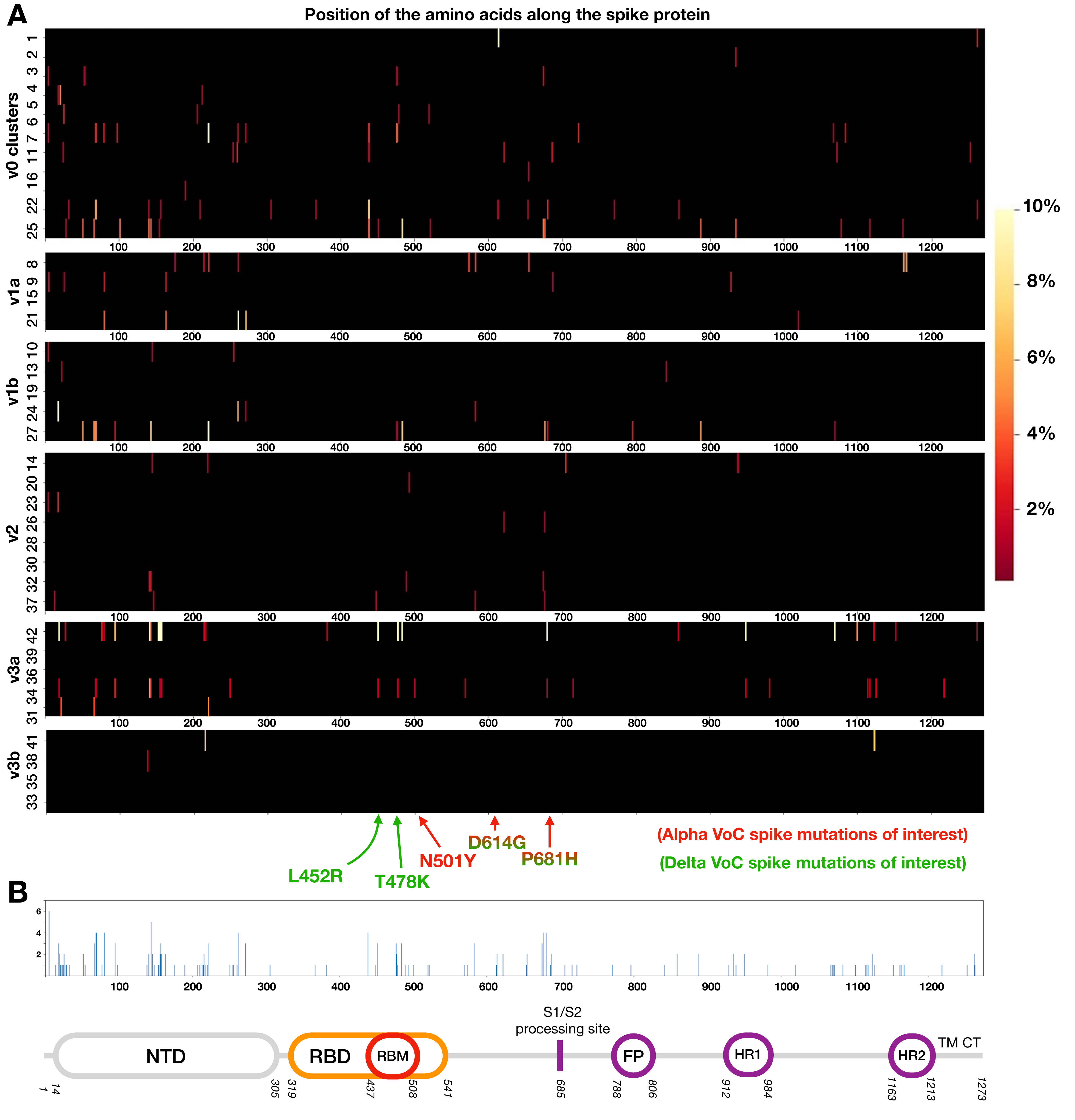}
\end{center}
\caption{{\bf Patterns of spike protein diversification of the emerging variants.} Heatmaps indicate for each cluster in a chain (A) the location and number of amino acid substitutions counted per all variants in a cluster when compared to the previous cluster and normalised accordingly (see supplementary material). (B) recapitulates the sum of all the values from all chains in A.}
\label{fig:heatmap}
\end{figure}

In Fig.~\ref{fig:heatmap} we also show the pattern of diversification occurring along the Spike protein in terms of single amino acid substitutions in the chains reconstructed with the cutoff $r_W=100$, i.e. v0, v1a, v1b, v2, v3a and v3b. The heatmaps in Fig.~\ref{fig:heatmap}A  associated to each main chain  show the position and the number of amino acid changes with respect to the previous cluster in the chain (see supplementary material for more details).
The picture shows where and how often a change occurs within the chain along time and offers an overview of these dynamics. 
Our time-ordered analysis along each chain allows to distinguish the extent of Spike mutations within stable chains. For instance, along v0 one can see higher variability in clusters 6 and 7, corresponding to when v1a/b and v2 branched off (see Fig.~\ref{fig:4}), and in the last two clusters before the variant disappeared from the data. Furthermore, we observe that v2 has a lower degree of variability compared to the other 5 chains. This could reflect the fact that v2 did not lead to new stable chains, while the other ones did, including those associated to the Delta VoC. This preliminary overview surely deserves further investigations, while it is here reported to highlight the additional level of details that a temporal variant analysis of the Spike protein offers.

As a final remark, our analysis also permits to observe the most conserved or variable regions per chain (regions were changes are minor or not occurring and regions with frequent substitutions). Interestingly, the comparison among the plots also shows that, although there are main hot spots of mutations that are detectable along the spike protein and along all the chains, each chain has a typical pattern of substitution.
Considered altogether (see Fig.~\ref{fig:heatmap}B), this analysis can allow us to identify regions of the Spike protein that also provide hints for more efficient targeting in monitoring or pharmaceutical interventions.

\subsubsection*{Epidemiological data and MeRG}

\begin{figure}[tb!]
\begin{center}
\includegraphics[width=.95 \textwidth]{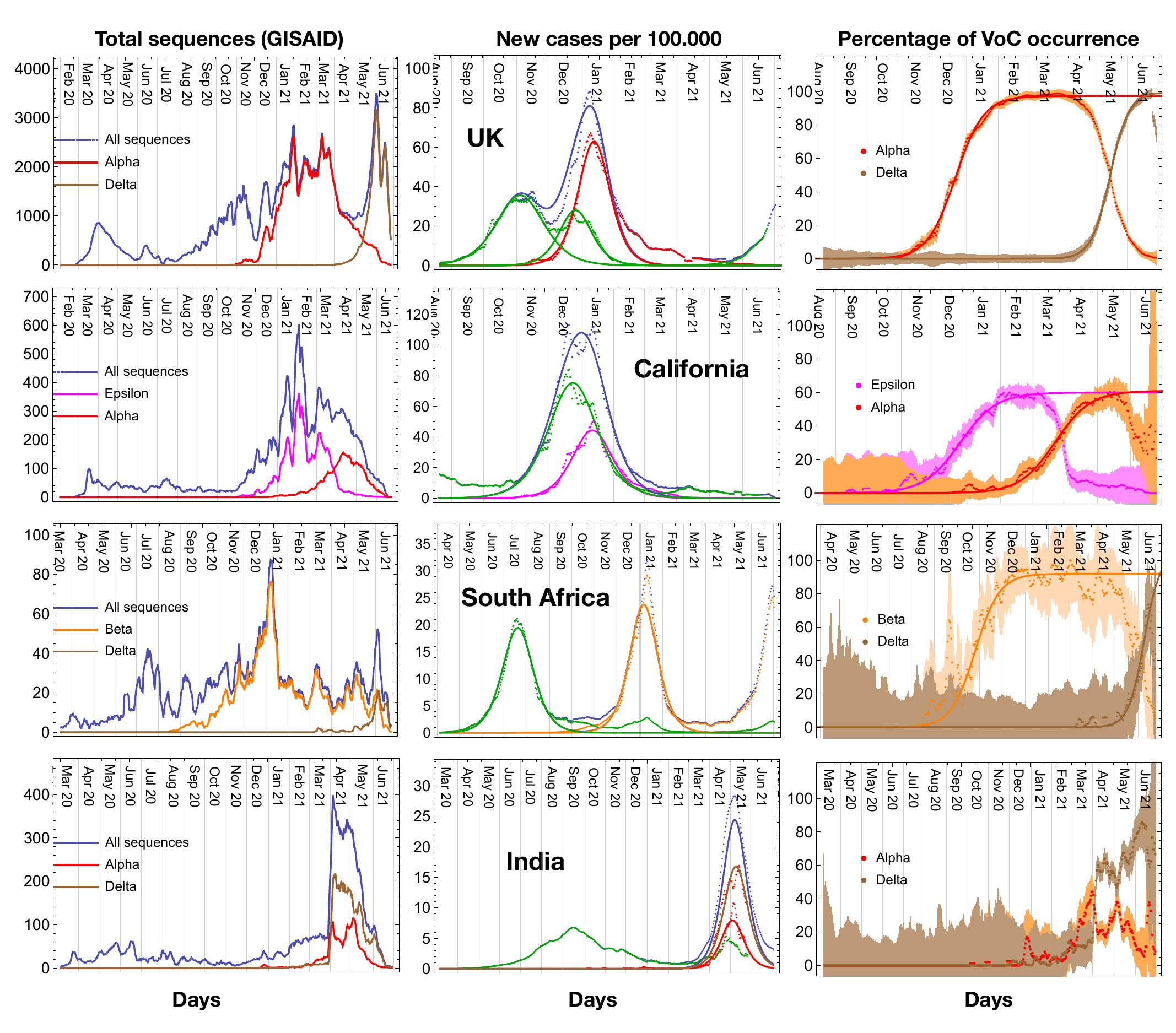}
\end{center}
\caption{{\bf MeRG model for epidemiological data of variants.} Results of the MeRG fitting of the number of infected associated to each relevant variant. Each row corresponds to a geographical region. In the left column we show the total number of sequencing available on GISAID (in colour the ones associated to the relevant VoC or VoI); the middle column shows the number of new daily infected (per 100.000 inhabitants); the right column shows the percentage of each VoC or VoI in the sequencing data. All plots show daily rates, with data smoothened over a period of 7 days. In the middle plots, the data are shown by dots, where blue corresponds to the total and the colours show the number of infected associated to each variant. The solid lines show the result of the fits to the MeRG model (note that only for the UK we fit the ``standard variant'' - in green - with two logistic functions). In the left plots, the error derives from the expected statistical variation on the number of daily sequences (after smoothening). For all the plots, the classification in variants derived from the GISAID data.}
\label{fig:5}
\end{figure}

The results of our ML analysis firmly suggest that there is a strong relation between the genesis of a new emerging variant and the onset of a new wave, with exponential increase in the number of infections, in the epidemiological data. In a companion article \cite{cacciapaglia2021MeRG} we developed a framework that can be used to describe the evolution of each variant. The model is based on the eRG approach by including mutations (MeRG). 

The MeRG framework models the time evolution of the cumulated number of infected by each variant in terms of a logistic function (sigmoid), solution of the eRG equation, and given by:
\begin{equation} \label{eq:1}
\mathcal{I}_c (t) = \frac{A e^{\gamma t}}{B + e^{\gamma t}}\,,
\end{equation}
where $\mathcal{I}_c $ is the cumulative number of infected, $\gamma$ is the infection rate (in inverse days) and $A$ is the total affected individuals after the wave (per 100.000 inhabitants). The parameter $B$ controls the timing of the wave, and is of no concern in this study. We recall that the parameter $\gamma$ encodes the effective diffusion speed of the variant, including not only its intrinsic viral power but also the effect of pharmaceutical measures (like vaccinations) and social distancing measures. Nevertheless, it is possible to compare the value of these parameters between different variants. If the diffusion occurs under similar social conditions, this represents a measure of the ability of the new variant to spread and infect new individuals.

Hence, we used the logistic function above to fit the epidemiological data, after distributing the new daily infected to each variant proportionally to the variant frequency observed in the sequencing data. This procedure yields a reliable estimate of the diffusion of each variant. For this purpose, we used the full dataset from GISAID for the whole UK, using the VoC classification embedded in the GISAID data. As shown before, this classification is equivalent to the result of our ML approach. The result is shown in the top row of Fig.~\ref{fig:5}, where we show the number of sequences (left plot), the new number of infections per variant and the result of the MeRG fit (middle) and the frequency of the VoCs (right). Note that the total numbers are plotted in blue, while the VoCs in colours. We considered the epidemiological data from the most recent waves, which developed between September 2020 and February 2021. The green curve in the middle plot shows that, after the first peak at the beginning of November, a second smaller peak developed. We describe the two with two independent sigmoids. The second sigmoid is subtracted from the data when fitting for the Alpha VoC data. The parameters from the fit are reported in Table\ref{tab:1}. As the social conditions during this period did not change substantially, it is meaningful to compare the $\gamma$ parameters for the Alpha and Delta VoC with the other ones (in green). We observed a marked increase in the infectivity, by 49\% for Alpha, which is compatible with laboratory tests.
Interestingly, the frequency percentage for the VoCs, shown in the left plot, can also be fitted very accurately with a logistic function in Eq.~\eqref{eq:1} as long as only one VoC dominates. The results are also reported in Table~\ref{tab:1}.  The fit parameter $\gamma_\%$ is a measure of how more infectious is the new VoC with respect to the previously dominant one. This plot also shows very effectively the switch between the two variants, occurring in May 2021.

\begin{table}[tb!]
\begin{center}
\begin{tabular}{|c||c|c||c|c|c||c||c|c|}
\hline
Region & \multicolumn{2}{c||}{standard variant} & \multicolumn{3}{c||}{variant of concern} & transmissibility & \multicolumn{2}{c|}{VoC percentage}\\ \hline
             &  $A$ & $\gamma$  &  $A_{\rm VoC}$ & $\gamma_{\rm VoC}$ &  VoC/VoI & increase & $A_{\%}$ & $\gamma_{\%}$ \\ \hline\hline
UK & $2140(12)$ & $0.0668(5)$ & $2530(10)$ & $0.0994(7)$ & Alpha & $49\%$ & $97.3(3)\%$ & $0.076(1)$ \\ 
 & $$ & $$ & $-$ & $-$ & Delta & $-$ & $99(1)\%$ & $0.115(2)$ \\ \hline
South Africa & $1104(2)$ & $0.0705(4)$ & $1161(2)$ & $0.0904(5)$ & Beta & $28\%$ & $91.9(8)\%$ & $0.061(4)$ \\
 & $$ & $$ & $-$ & $-$ & Delta & $-$ & $96(6)\%$ & $0.090(7)$ \\ \hline
India & $717(3)$ & $0.0358(4)$ & $497.8(8)$ & $0.0858(3)$ & Alpha & $140\%$ & $-$ & $-$ \\
 &  &  & $908(5)$  & $0.0747(6)$ & Delta & $109\%$ & $-$ & $-$ \\ \hline
California & $4773(7)$ & $0.0620(3)$ & $2250(5)$ & $0.0758(5)$ & Epsilon & $22\%$ & $59.9(6)\%$ & $0.059(4)$ \\
 & $$ & $$ & $-$ & $-$ & Alpha & $-$ & $61.0(6)\%$ & $0.0610(2)$ \\ \hline
\end{tabular}
\end{center}
\caption{{\bf MeRG fit parameters.} Parameters from the fit of the VoC/VoI for the UK, South Africa, California and India, also shown in Fig.~\ref{fig:5}.  The fit follows the MeRG model, according to which each variant can be fitted by an independent logistic function. For the UK, the ``standard variant'' fit corresponds to the first peak, in October-November 2020. The transmissibility increase is computed by comparing the gamma of the VoC with that of the standard variant in the same country. For the new variants that have not reached the peak of diffusion, it is not possible to 
extract reliable values for the eRG parameters.\label{tab:1}}
\end{table}

We repeated the same analysis for South Africa, California and India, which show very good fits notwithstanding the more limited sequencing statistics available on GISAID. This is clearly shown in the left plots, where we report the statistical uncertainly at 65\% confidence level, due to the available sequencing.  The results, shown in Fig.~\ref{fig:5} and Table~\ref{tab:1}, demonstrate that the MeRG framework provides an excellent modelling of the data.

\subsubsection*{Performance of the ML algorithm as an early warning tool for emerging variants}

The results for time-ordered chains with monthly clustering (Fig.~\ref{fig:4}) demonstrate that our ML algorithm is able to efficiently identify the emergence of new variants that have strong impact on the epidemiological evolution of the disease. Hence, it can be used as an early warning tool for new potentially dangerous variants that may become VoCs. To test the performance of this tool, we validated the procedure on the emergence of the Alpha VoC.
The first Alpha VoC case has been found on the 20th of September, 2020. Following the monthly analysis, we identified a cluster dominated by the Alpha VoC in November 2020, at the beginning of chain v2. Once the first cluster is identified, one would need to add the data for the following months to confirm its persistence (as the process is additive, the cluster definitions in the previous months are not affected). 
We saw empirically from Fig.~\ref{fig:4} that persisting chains contain at least three clusters, hence an emerging variant could be defined only two months later (January 2021 for the Alpha VoC).

\begin{table}[ht]
\centering
\begin{tabular}{|c|c|c|c|c|l|}
\hline
Cal. week & Date (Mon) & Total seq. &  \% of Alpha VoC & n$^{\rm o}$ of clusters & \\
\hline
38 & 14 Sept. & $1948$ & $0.05$ & -- &  First detection \\ \hline
39 & 21 Sept. & $3394$ & $0.06$ & -- &  \\\hline
40 & 28 Sept. & $2203$ & $0.09$ & 5 &  \\\hline
41 & 5 Oct. & $3891$ & $0.1$  & 5 & \\\hline
42 & 12 Oct. & $4598$ & $0.07$ & 5 &  \\\hline
43 & 19 Oct. & $5921$ & $0.4$ & 2 &  \\\hline
44 & 26 Oct. & $4557$ & $1.5$ & 1 & First warning (weekly) \\\hline
45 & 2 Nov. & $7589$ & $3$ & 1 &  \\\hline
46 & 9 Nov. & $7200$ & $7$ & 1 &  \\\hline
47 & 16 Nov. & $4669$ & $12$ & 1 &  Emerging persistent variant (weekly)\\\hline
48 & 23 Nov. & $2343$ & $12$ & 1 &  \\\hline
49 & 30 Nov.. & $1971$ & $21$ & 1 & First warning (monthly)\\\hline
50 & 7 Dec. & $6382$ & $38$ & 1 & \\\hline
51 & 14 Dec. & $8059$ & $50$ & 1 & \\\hline
52 & 21 Dec. & $4864$ & $53$ & 1 &  \\\hline
53 & 28 Dec. & $7766$ & $65$ & 1 & WHO classification as VoC \\\hline
\end{tabular}
\caption{\label{tab:weekly_dataset} {\bf Early warning for the Alpha VoC.} Results of the ML analysis applied to weekly binned data after the first detection of the Alpha VoC in England. The columns contain the 2020 calendar week number, with the initial date (Monday), the total number of sequences in the dataset for each week and the percentage of Alpha VoC sequences identified a posteriori in the data. The fifth column contains the minimal number of clusters in the ML output that allows to isolate the Alpha VoC cases. When this indicator is equal to 1, an early warning can be issued (week 44). After three weeks of the cluster persisting, we identify it with an emerging variant (week 47). This date can be compared to the WHO classification decision (29 Dec., week 53).}
\end{table}

\begin{figure}[tb!]
\begin{center}
\includegraphics[width=.85 \textwidth]{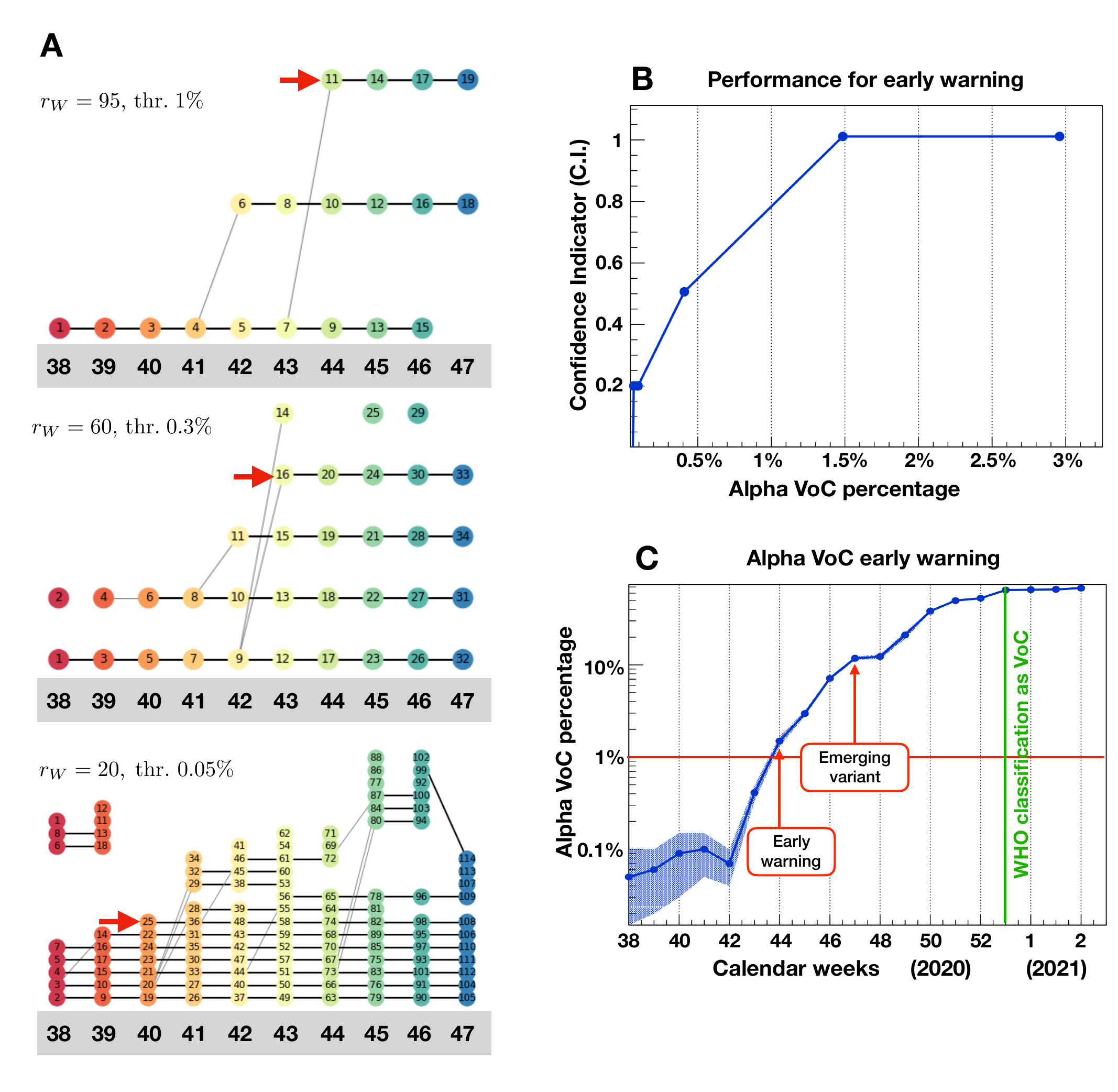}
\end{center}
\caption{{\bf Early warning performance.} A) Cluster chains from week 38 to week 47 obtained at different working points. The chain containing the Alpha VoC Spike variant is highlighted by the red arrow. B) Confidence indicator for the early warning performance as a function of the percentage of the new variant in the time-binned data. C) Data for the Alpha VoC, indicating the early warning and definition of the emerging variant, using our ML on a weekly binning, compared to the time of the WHO classification as VoC.}
\label{fig:EW}
\end{figure}

To improve the performance of the ML algorithm in terms of prediction, we reanalysed the same data using a weekly- rather than monthly-based binning (Table~\ref{tab:weekly_dataset}). The cutoff on the Ward distance needs to be adjusted for the weekly analysis, thus leading to a choice that differs slightly from that of Fig.~\ref{fig:4}.
Using a cutoff $r_W = 95$ (instead of $100$) as well as the 1\% threshold, the Alpha VoC chain was identified as branching off in week 44, at a moment when this variant represented $1.5\%$ of the total weekly sequences, as illustrated by the top diagram in Fig.~\ref{fig:EW}A. By week 46 (stable chain of at least three clusters) this analysis would have been able to identify the Alpha variant as VoC by the 9th of November.
We next analysed the same data by lowering both cutoff and threshold to determine whether this could influence the performance of the ML algorithm.
Using less stringent parameters it is possible to define separate clusters containing the Alpha VoC sequences earlier than week 44 (Fig.~\ref{fig:EW}A lower two panels, highlighted by a red arrow). However, in these cases the analysis yields to an increase in the number of clusters and chains, which makes it difficult to identify unequivocally new emergent variants.  To quantify the performance, therefore, we counted how many additional chains appear when an Alpha VoC chain can be isolated at each week, as shown in the fifth column of Table~\ref{tab:weekly_dataset}. For all weeks after 44 included, it suffices to generate one new cluster besides the two that contain the dominant Spike variant of v0 and v1, while for week 43 at least two new clusters are necessary.
The inverse of the number of clusters defined above quantifies a ``confidence indicator'' (C.I.) for the early warning, and it is plotted in Fig.~\ref{fig:EW}B as a function of the Alpha VoC percentage in each week dataset. The C.I. can be interpreted as the probability of identification of the correct emerging variant via the first cluster. The result shows that probabilities above 50\% require the presence of the new variant in at least 1\% of the sequences. In Fig.~\ref{fig:EW}C, we show the Alpha VoC percentage as function of time in weeks. An early warning can be issued as soon as the new variant surpasses 1\% of the data, leading to an early warning for the Alpha VoC in week 44, i.e. 6 weeks after the first detection. If the chain persists for 3 more weeks, an emerging variant is identified in week 47, hence 6 weeks before the official classification as a VoC by WHO.
It is possible to reduce the time-scale below a week by increasing the sequencing. To obtain a result with statistical uncertainty below 10\%, for instance, a few thousand sequences in each time bin would be required. For the Alpha VoC data, this is achieved for weekly binning, as shown in Table~\ref{tab:weekly_dataset} (the statistical error is shown as a band in Fig.~\ref{fig:EW}C). 

At the moment of submission of this work, there is an increasing  coverage in the media about the possible concerns due to the raise of a new variant in the UK (AY.4.2 Pango lineage). 
Our monthly analysis already identified this lineage as a dominant Spike variant in cluster 43 (Fig.~\ref{fig:4}), branching off from the main Delta chain in August 2021 (date at which data acquisition was stopped for the purposes of this work). To determine whether this novel variant could be considered of concern, we carried out our ML analysis with a weekly binning. Despite the fact that this analysis was carried out with a suboptimal $r_w =100$ (due to the urgency of the situation analyses to determine the optimal $r_w$ are ongoing), our weekly analysis clearly indicates that AY.4.2 Pango lineage has formed a stable chain of 3 clusters by the19th of September, 2021. This analysis thus indicates that this variant is truly establishing in the UK as a variant of concern.

\section*{Discussion}

We presented a ML algorithm that allows to identify, classify and track epidemiologically relevant variants of SARS-CoV-2. It is based on the Levenshtein distance of the Spike protein sequences and is unbiased in the sense that it requires no prior knowledge of any of the variants' properties. For each time bin, the algorithm first produces an independent clustering of the Spike protein sequences. It then links clusters in subsequent bins with a common dominant Spike variant, thereby creating chains of clusters depicting temporal waves of variants.  The results for England empirically showed that the a chain persisting at leat 3 consecutive clusters is a strong indication for an increased viral fitness of its dominant variant. This criterion allowed to identify emerging variants that pose a significant epidemiological threat. We validated the method with both monthly and weekly time binning.

We applied the algorithm to the sequencing data from England, which offers the largest dataset on the GISAID open-source genome repository. Among the emerging variants, the officially recognised VoCs (Alpha and Delta) were clearly identified and isolated. Similar results for Wales and Scotland (despite a more limited number of available sequences) confirmed the effectiveness of the algorithm, while comparison of our approach (that uses data of the Spike protein sequences only) to other informed methods based on the complete genome validated the algorithm. Furthermore, the temporal organisation of clusters into chains served as a tool not only to monitor the genetic evolution of the Spike protein but also to help shed light on its mechanisms. On the one hand, from the temporal chain analysis branching relations arouse from which we can reconstruct the evolutionary diversification that leads to the establishing of emergent variants. On the other hand, within a single chain, the analysis of mutations of subdominant Spike variants permitted to distinguish regions of the sequence with a high frequency of mutations from those in which no amino acid substitutions take place over time. 

Using the relative percentage of each variant in the sequencing dataset to estimate the number of individuals infected by each variant, we correlated our temporal chain analysis with epidemiological data. We discovered that each new wave of the COVID-19 pandemic in England (and similarly in Scotland and Wales) was driven and dominated by a new emerging variant. This observation corroborates the hypothesis that there exists a strong and direct causal relation between the emergence of a new variant and the onset of a new epidemic wave. We modelled the cumulative number of infected individuals by use of the MeRG framework that we proposed in a companion manuscript \cite{cacciapaglia2021MeRG}. We also used epidemiological data from the whole UK, California, India and South Africa to confirm the validity of the model.

Finally, in view of potential future waves of COVID-19, we tested the viability and performance of our ML algorithm as an early warning tool to detect the emergence of a new, epidemiologically dangerous, variant. We demonstrated that the Alpha VoC could be established as the dominant variant of an emerging persistent chain 9 weeks after its first detection in the England data set. This precedes its classification as a VoC by the WHO by 6 weeks. We showed, more generally, that an early warning for the emergence of a new persistent variant can be issued once its associated cluster reaches 1\% of the time-binned sequence data. Interestingly, the Spike protein was a reliable and sufficient reference to meet a successful goal.
Despite being preliminary in light of the need to better adjust the clustering cutoff, our analysis of the most recent data in the UK stresses the emergence of the  Pango lineage AY.4.2 into a stable chain and thus of a true variant of concern. Our analysis indicates that an early warning could have been issued as of the 19th of September. Thus, our ML analysis tool and these early warning indications could be used by policy makers to  implement  immediate actions  that globally limit the spread of this and of other variants that will emerge in the future.

\subsection*{Limitations}

This study was performed on the sequencing data from a single region, England. This is justified by the fact that the sequencing dataset associated to England on the GISAID open-source genome repository is by far the largest and most uniform compared to other countries/regions. Biases relative to specific data-taking practices un England may induce biases in the analysis.
To validate the results, however, we have also analysed the data for Wales and Scotland, as presented in the supplementary material. We chose the other two nations of the Great Britain island because they have a very similar epidemiological history compared to England, thus we would expect comparable outcomes. As such, by comparing the results we would test the reliability of the ML procedure alone. In fact, the results for Wales and Scotland, while less significant with respect to statistics, show the same patterns we obtained for England. 

As an early warning tool, our ML approach is triggered when a new cluster chain branches off. This procedure is sensitive to the working point chosen for the clustering algorithm, and it has a certain chance to produce a false positive. However, the extensive use of this tool on future data and on sequences from other countries/regions will allow to estimate more reliably the false positive probability.

\subsection*{Conclusions}

The results of our ML analysis have profound impact, both scientifically and epidemiologically: They provide new insights that are crucial for the development of new strategies to study how SARS-CoV-2 variants emerge and to predict the evolutionary pattern as well as the characteristics of future mutations of the Spike protein. We provide a tool that allows for an efficient and unbiased identification of emerging variants, for the tracking of the evolution and diversification of their Spike proteins and, most importantly, for an early warning system to identify epidemiological threats for the population.

The concrete results presented in this work constitute a milestone for the development of a new exploratory strategy of the genesis of variants in epidemic or pandemic infectious diseases. The temporal dynamics of variants, in fact, allows to study the branching off of new relevant variants, and to track the evolutionary pattern of amino acid substitutions in the Spike protein highlighting persistent variants and structural trends, in terms of hotspots of mutations and/or of conserved regions. Further studies are necessary to fully exploit this information. 

Our work furthermore underlines the importance of sufficient genomic data in order to both scientifically understand and track the temporal evolution of viral diseases, but also to issue sufficiently early warnings of epidemiologically dangerous variants so that decision makers can take efficient preventative measures. Our approach can be applied to other viral diseases, like influenza, provided that sufficient sequencing data is available. 

\section*{Acknowledgements}
We acknowledge with gratitude the authors, originating and submitting laboratories of the genetic sequence and metadata made available through GISAID. A full listing of all authors and laboratories is available on the GISAID website.

\section*{Author contributions statement}

This work has been designed and performed conjointly and equally by all the authors. In particular: AdH, SV, AG and FC have developed the Machine Learning algorithm and analysed the spike protein sequencing data; AC and MLC contributed in the analysis of the spike protein diversification; CC has collected and analysed the clade and variant of concern data from GISAID; GC, CC, SH and FS have developed the theoretical framework. All authors have equally contributed to the writing of the text.

\section*{Additional information}

The authors declare no competing interests.

\section*{Data and code availability}

All raw data used in this work are obtained from open-source repositories: \href{https://www.gisaid.org/}{GISAID} for the sequencing; \href{https://ourworldindata.org/}{Ourworldindata.org} and the \href{https://coronavirus.data.gov.uk/details/cases}{UK Coronavirus Dashboard} for the epidemiological data. The Machine Learning code is available at \href{https://github.com/AdeledeHoffer/ML-Covid}{https://github.com/AdeledeHoffer/ML-Covid}.

\bibliography{biblio}

\end{document}